\documentclass[twocolumn,showpacs,preprintnumbers,amsmath,amssymb]{revtex4}
\usepackage[dvips]{graphicx} 
\begin{document}
\preprint{IMAFF-RCA-05-06}
\title{Coherent states in quantum cosmology}

\author{S. Robles-P\'{e}rez$^{1,2}$, Y. Hassouni$^{3}$ and P. F.
Gonz\'{a}lez-D\'{i}az$^{1,2}$} \affiliation{$^{1}$Colina de los
Chopos, Centro de F\'{\i}sica ``Miguel Catal\'{a}n'', Instituto de
Matem\'{a}ticas y F\'{\i}sica Fundamental,\\ Consejo Superior de
Investigaciones Cient\'{\i}ficas, Serrano 121, 28006 Madrid
(SPAIN). \\$^2$Estaci\'{o}n Ecol\'{o}gica de Biocosmolog\'{\i}a,
Medell\'{\i}n (SPAIN).
\\$^{3}$Laboratoire de Physique Theorique, Faculte des
sciences-Universite Mohamed V-Agdal, \\ Avenue Ibn Batouta B.P:
1014, Agdal Rabat (MOROCCO). }
\date{\today}
\begin{abstract}
In the realm of a quantum cosmological model for dark energy in
which we have been able to construct a well-defined Hilbert space,
a consistent coherent state representation has been formulated
that may describe the quantum state of the universe and has a
well-behaved semiclassical limit.
\end{abstract}


\maketitle

\section{Introduction}

In a previous paper \cite{GonzalezDiaz07} a general, simple
quantum description was constructed for a model of an homogeneous
and isotropic universe filled with a fluid described by an
equation of state, $p=w \rho$, being $p$ and $\rho$ the pressure
and the energy density of the fluid, respectively, and $w$ is a
constant parameter. That model can be regarded to be an
approximate idealization in which only a particular kind of energy
dominates the universe along its time evolution, from the
beginning to the end. Among these particular dominating energies,
most emphasis was paid to the case of dark energy. The great
advantage of this mode is that it is analytically solvable and,
therefore, able to neatly show the analogy between quantum
mechanical open systems and quantum cosmology, and take it quite
far at least formally, even though quantum cosmology adds some
exceptional features to the formalism. Thus, an analytically
solvable model for which the complete quantum development can be
tracked can be considered.

On the other hand, coherent states have been always considered as
mathematical objects with applications in quantum physics. In
fact, the large number of their applications lead to the
introduction for new definitions of particular quantum systems for
which coherent states are involved. Coherent states can be
constructed from the algebras which are behind their definition.
More precisely, in the literature we usually deal with Heisenberg
algebras to obtain them. Nevertheless, in some works
\cite{Klauder63,Quesne02} coherent states for some quantum systems
are constructed from the so-called Generalized Heisenberg Algebras
(GHA).

In this paper we shall take advantage from the above property of
the model to investigate the role that coherent states, obtained
from a GHA, may play in a cosmological model. We in fact obtain
general expressions for the cosmic wavefunctions that describe
coherent states, which can be taken as a basis for further
developments of this subject.

We outline the paper as follows. In sec. II we give a brief
summary of the cosmological model, reviewing the basics aspects of
its Hilbert space. Coherent states are formulated and described in
sec. III and, in section IV, we give some conclusions and further
comments.

\section{A cosmological model}

The model considered in ref. \cite{GonzalezDiaz07} consists of a
Friedman-Lemaître-Robertson-Walker (FLRW) universe filled with a
fluid described by the equation of state, $p=w \rho$, where $w$ is
a constant parameter. For a gauge $\mathcal{N}=a^3$, where
$\mathcal{N}$ is the lapse function and $a\equiv a(t)$ is the
cosmic scale factor, the Hamiltonian of the system is given by,
\begin{equation}
H = - \frac{2 \pi G}{3} a^2 p^2_a + \rho_0 a^{3(1-w)} ,
\end{equation}
where $p_a$ is the conjugate momenta to the scale factor, $G$ is
the gravitational constant, and $\rho_0$ is the energy density of
the fluid at the coincidence time \cite{GonzalezDiaz07}. Then, a
set of Hamiltonian eigenfunctionals can be obtained. In the
configuration space, they are given as
\begin{equation}\label{Hamiltonian eigenfunctionals}
\phi_n(a) = N_n a^{\alpha} \mathcal{J}_n(\lambda a^q) ,
\end{equation}
in which $N_n$ is a normalization constant, $\alpha$ is a
parameter depicting the factor ordering ambiguity, $\mathcal{J}_n$
is a Bessel function of the first kind and order $n$, and,
\begin{equation}
q = \frac{3}{2}(1-w) \, \, , \, \, \lambda = \frac{1}{\hbar q}
\sqrt{\frac{3}{2 \pi G} \rho_0}.
\end{equation}
The functions given by Eq. (\ref{Hamiltonian eigenfunctionals})
correspond to the following eigenvalue problem,
\begin{equation}
\hat{H} \phi_n(a) = \mu_n \phi_n(a) ,
\end{equation}
with,
\begin{equation}\label{Hamiltonian eigenvalues}
\mu_n = q^2 n^2 - \epsilon_0^2 ,
\end{equation}
where $\epsilon_0^2$ is a factor which depends on the factor
ordering choice. Now, we have to impose some boundary conditions
in order to construct wavefunctionals which can describe the state
of the universe. In particular, when the fluid is dark energy ($w
< -\frac{1}{3}$)\cite{GonzalezDiaz07}, the boundary conditions
are: i) the wavefunctionals have to be regular everywhere, even
when the metric degenerates, $a\rightarrow 0$, and ii) they should
vanish at the big rip singularity when $a\rightarrow \infty$. The
boundary conditions are satisfied by the Hamiltonian
eigenfunctionals when we impose the following restrictions on the
parameter $\alpha$,
\begin{equation}
- q n \leq \alpha < \frac{q}{2} \sim \frac{3}{2} .
\end{equation}
Now, in order to develop the usual machinery of quantum mechanics,
we need to construct a well-defined Hilbert space. It is usually
an impossible task, in general, when gravitational fields are
taken into account, since they appear non-renormalizable
infinities in the formalism. Just in the case of very simplified
minisuperspaces, a regularization process can be made and, then, a
Hilbert space can be however defined.

We can start by defining the Hamiltonian eigenstates, $|n>$, to be
those states represented in the configuration space by the
wavefunctionals given in Eq. (\ref{Hamiltonian eigenfunctionals}),
i.e.,
\begin{equation}
<n|a> = <a|n> = \phi_n(a) ,
\end{equation}
as the wave functionals considered so far are real functions.
Then, let us define the scalar product to be,
\begin{equation}\label{scalar product general}
<f|g> = \int_0^{\infty} da \, W(a) f(a) g(a) ,
\end{equation}
where $W(a)=a^k$ is a weight function.Thus, the orthogonality
relations between the Hamiltonian eigenstates turn out to be,
\begin{equation}\label{orthogonality relations 0}
<n|m> = \frac{N_n N_m}{q} \int_0^{\infty} du \, u^{\frac{k+2\alpha
+1}{q}-1} \mathcal{J}_n(\lambda u) \mathcal{J}_m(\lambda u) ,
\end{equation}
with $u=a^q$, and using the standard bibliography
\cite{Gradshteyn03}, this integral can be performed for the
following values,
\begin{equation}
n+m+1 > 1 -\frac{k+2\alpha +1}{q} >0 .
\end{equation}
For instance, for a weight function, $W(a) =
a^{\frac{q}{2}-(2\alpha +1)}$, the orthogonality relations are,
\begin{equation}\label{orthogonality relations 1}
<n|m> = \sqrt{\frac{\pi}{2 q \tilde{\lambda}}} \frac{N_n N_m \,
\Gamma(\frac{1}{2})
\Gamma(\frac{2(n+m)+1}{4})}{\Gamma(\frac{2(m-n)+3}{4})
\Gamma(\frac{2(n+m)+3}{4}) \Gamma(\frac{2(n-m)+3}{4})} ,
\end{equation}
where, $\tilde{\lambda} = q \lambda =
\frac{1}{\hbar}\sqrt{\frac{3}{2 \pi G} \rho_0}$. In particular,
they do not show any problem with the normalization of the zero
mode because,
\begin{equation}
\langle 0 | 0 \rangle = N_0^2 \sqrt{\frac{\pi}{2 q
\tilde{\lambda}}}
\frac{\Gamma(\frac{1}{4})}{\Gamma(\frac{3}{4})^3} ,
\end{equation}
which can be normalized by choosing an appropriate value of the
normalization constant, $N_0$. However, by Eq. (\ref{orthogonality
relations 1}), these relations do not form an orthogonal basis for
the representation of the quantum state of the universe.
Nevertheless, using the scalar product (\ref{scalar product
general}), we can define an orthonormal basis, in terms of
Laguerre polynomials. For the value $k=\frac{q}{2}-(2 \alpha +1)$
in the integration measure, we can use the following set of
functionals,
\begin{equation}
\psi_n(a) = N_n a^{\frac{4\alpha+q}{4}} e^{-\frac{\lambda a^q}{2}}
L_n(\lambda a^q) ,
\end{equation}
where,
\begin{equation}
L_n(x) = \sum_{m=0}^n \left( \begin{array}{c} n
\\ m\end{array} \right)  \frac{(-x)^m}{m!} ,
\end{equation}
is the Laguerre polynomial of order $n$. They form an orthonormal
basis with appropriate values of the normalization constants,
$N_n$. Then, the Hamiltonian eigenstates can be decomposed into
the basis defined by the set $\{\psi_n \}$ as,
\begin{equation}
\phi_n(a) = \sum_{m=0}^{\infty} C_{mn} \psi_m(a),
\end{equation}
where the coefficients, $C_{nm}$, are given by,
\begin{equation}
C_{nm} = \langle \psi_m | \phi_n \rangle = \int_0^{\infty} da \,
a^{\frac{q}{2}-(2\alpha + 1)} \psi_m(a) \phi_n(a) .
\end{equation}
We can then develop the usual formalism of quantum mechanics in
this orthonormal basis.

The standard procedure of constructing coherent states is clearer
when we work with an orthonormal basis of Hamiltonian eigenstates.
Let us therefore consider the scalar product (\ref{scalar product
general}), for $k=-(2\alpha + 1)$. In that case, the orthogonality
relations for the Hamiltonian eigenstates can be written as
\cite{GonzalezDiaz07},
\begin{equation}\label{orthogonality}
\begin{array}{ll}
\langle n | n \rangle = 1 &, \forall n, \\ & \\ \langle n | m
\rangle = 0 & , |n - m| \; \; \;
\textmd{even,} \\ & \\
\langle n | m \rangle = \frac{4}{\pi}
\frac{(-1)^{\frac{1}{2}(n-m-1)} \sqrt{n \, m}}{n^2 - m^2} & , |n -
m|  \; \; \; \textmd{odd}.
\end{array}
\end{equation}
The price that we have to pay when using this scalar product is
that then we need a regularization procedure for the zero mode
\cite{GonzalezDiaz07}. Its advantage is that the set of
Hamiltonian eigenstates can be split in two sets, for even and
odds modes, which are orthogonal to each other, separately. Then,
we can use an analogous formalism to that developed in
\cite{Gazeau03}. Let us split the Hilbert space as
\begin{equation}
\mathcal{H} = \mathcal{H}_+ \oplus \mathcal{H}_- ,
\end{equation}
where the subspaces, $\mathcal{H}_+$ and $\mathcal{H}_-$, are
chosen to be,
\begin{eqnarray}
f_+(u) \in \mathcal{H}_+ & \Rightarrow f_+(u) & =
\sum_{n=0}^{\infty} C_{2n} \phi_{2n}(u) \otimes \chi_+ \\ f_-(u)
\in \mathcal{H}_- & \Rightarrow f_-(u) & = \sum_{n=0}^{\infty}
C_{2n+1} \phi_{2n+1}(u) \otimes \chi_- ,
\end{eqnarray}
with some constants $C_k$ , and,
\begin{equation}
\chi_+ = \left( \begin{array}{c} 1
\\ 0 \end{array}\right) \; \;\; , \; \;\; \chi_- = \left( \begin{array}{c}
0 \\ 1 \end{array}\right) .
\end{equation}
The subspaces, $\mathcal{H}_+$ and $\mathcal{H}_-$, turn out to be
the subspaces of even and odd functionals, and any function
belonging to the space $\mathcal{H}$ can be decomposed as,
\begin{equation}
f(u) = f_+(u) \oplus f_-(u) .
\end{equation}
Let us define now, in this space, the following scalar product for
any two functions, $f(u),g(u) \in \mathcal{H}$,
\begin{widetext}
\begin{equation}
\langle f | g \rangle = \lim_{l_p \rightarrow 0}
\int_{l_p}^{\infty} du \, W_1(u) f^{\dagger}(u) g(u) = \lim_{l_p
\rightarrow 0} \int_{l_p}^{\infty} du \, W_1(u) \left( f_+(u)
g_+(u) + f_-(u) g_-(u) \right) ,
\end{equation}
\end{widetext}
weighted by the function, $W_1(u) = u^{-(2 \alpha + 1)}$, with the
limit being introduced to regularize the zero mode. With this
scalar product the basis $\left\{ \phi_n \right\}_{n \in N}$
becomes orthonormal,
\begin{equation}
\langle n | m \rangle = \langle \phi_n | \phi_m \rangle = \xi
\lim_{l_p \rightarrow 0} \int_{l_p}^{\infty} du \, \frac{1}{u}
\mathcal{J}_n(u) \mathcal{J}_m(u) ,
\end{equation}
where $\xi$ is given by the usual scalar product between the
vectors, $\chi_{\pm}$, i.e.,
\begin{eqnarray}
 \begin{array}{ccc} n,m \;\; \textrm{even} & \Rightarrow & \xi =
\chi_+^{\dagger} \chi_+ = 1 \\ n,m \;\; \textrm{odd} & \Rightarrow
& \xi = \chi_-^{\dagger} \chi_- = 1 \\ n \;\; \textrm{even}, m
\;\; \textrm{odd} & \Rightarrow & \xi = \chi_+^{\dagger} \chi_- =
0 \\ n \;\; \textrm{odd}, m \;\; \textrm{even} & \Rightarrow & \xi
= \chi_-^{\dagger} \chi_+ = 0 .
\end{array}
\end{eqnarray}
Therefore, we have obtained an orthogonal basis of Hamiltonian
eigenfunctionals for a universe filled with dark energy. Now, we
can apply the formalism described in ref. \cite{Hassouni04} to
construct coherent states for the model being considered.

\section{Coherent states}

The interest of formulating coherent states in cosmology is two
fold. On the one hand, this would prepare the mechanics of the
universe to further, potentially, generalizable new developments,
and, on the other hand, to enhance the analogy between usual
quantum mechanics and cosmology.

In what follows we shall use the formalism to construct coherent
states for a system described by a generalized algebra
\cite{Hassouni04}, given by
\begin{eqnarray}
H_0 A^{\dag} & = & A^{\dag} f(H_0) \\ A H_0 & = & f(H_0) A \\
\left[ A^{\dag}, A \right] & = & H_0 - f(H_0) ,
\end{eqnarray}
where $A$, $A^{\dag}$ and $H_0$ are the generators of the algebra,
and $f(x)$ is called the characteristic function of the system.
$H_0$ is the Hamiltonian of the physical system under
consideration, with eigenstates given by
\begin{equation}
H_0 |m \rangle = \epsilon_m |m\rangle ,
\end{equation}
and $A^{\dag}$ and $A$ are the generalized creation and
annihilation operators,
\begin{eqnarray}
A^{\dag} |m\rangle &=& N_m |m+1\rangle \\ A|m\rangle &=& N_{m-1}
|m-1\rangle ,
\end{eqnarray}
where, $N_m^2 = \epsilon_{m+1} -\epsilon_0$. The use of a
generalized algebra \cite{Hassouni04} would add a parametrization
through the characteristic function, $f(H_0)$, that allows us to
have a systematic covering of distinct potentials for the given
system. The customary Heisenberg algebra is recovered in the limit
value $f(x) = 1+x$ \cite{Hassouni04}.

Then, the coherent states are defined to be the eigenstates of the
generalized annihilation operator,
\begin{equation}
A|z\rangle = z |z\rangle ,
\end{equation}
where $z$ is a generally complex number.

Since we have a Hamiltonian spectrum for the model of a dark
energy dominated universe, $H |n\rangle = \epsilon_n |n\rangle$,
we can now find the characteristic function, $f(x)$, and the
quantum excitation levels can be written as $ \epsilon_{n+1} =
f(\epsilon_n)$ \cite{Curado01}. Choosing a factor ordering
$\alpha=\beta$ so that $\epsilon_0^2=0$, we have,
\begin{equation}
\epsilon_{n+1} = q^2 (n+1)^2 = \epsilon_n + 2 q \sqrt{\epsilon_n}
+ q^2 = (\sqrt{\epsilon_n} + q)^2 = f(\epsilon_n) .
\end{equation}
The spectrum of the case being considered is formally similar to
the spectrum for a free particle in a square well potential
\cite{Hassouni04}, and the computation to follow can be done in a
parallel way.

Thus, the coherent states are given by,
\begin{equation}
|z\rangle = N(z) \sum_{n=0}^{\infty} \frac{z^n}{N_{n-1}!}
|n\rangle ,
\end{equation}
where,
\begin{equation}
N_n! \equiv N_0 N_1 \cdots N_n ,
\end{equation}
with, for consistency, $N_{-1}! \equiv 1$. Therefore, since
$N_{n-1}^2 = \epsilon_n - \epsilon_0$, it can be checked that in
our case,
\begin{equation}
N_{n-1}! = q^n n! ,
\end{equation}
and the coherent states can be written then as,
\begin{equation}
|z\rangle = N(z) \sum_{n=0}^{\infty} \frac{z^n}{q^n n!} |n\rangle
.
\end{equation}
This expression can be formally simplified, in terms of the
creation operator, since the state $|n\rangle$ can be written as,
\begin{equation}
|n\rangle = \frac{1}{N_{n-1}!} (A^{\dag})^n |0\rangle =
\frac{1}{q^n n!} (A^{\dag})^n |0\rangle.
\end{equation}
The coherent states can be expressed then with a formal
expression,
\begin{equation}\label{coherent states creation operator}
|z\rangle = N(z) \sum_{n=0}^{\infty} \left( \frac{z A^{\dag}}{q^2}
\right)^n \frac{1}{(n!)^2} |0\rangle = N(z) I_0\left( 2
\sqrt{\frac{z A^{\dag}}{q^2}} \right) |0\rangle ,
\end{equation}
where, $I_0(x)$, is the modified Bessel function of the first kind
of order zero.

Now, we have to impose the following conditions in order to get
the so-called Klauder's coherent states \cite{Hassouni04} (KCS):

i/ Normalization:
\begin{equation}
\langle z | z \rangle = 1 ,
\end{equation}

ii/ Continuity in the label, $z$:
\begin{equation}
|z-z'| \rightarrow 0 \;\; \Rightarrow \;\; || \, |z\rangle -
|z'\rangle \, || \rightarrow 0 ,
\end{equation}

iii/ Completeness:
\begin{equation}
\int d^2z \; W(z) \, |z\rangle \langle z| = 1 .
\end{equation}

The normalization condition can be found by choosing an
appropriate normalization function, $N(z)$. In terms of the
Hamiltonian eigenvectors, the norm of the coherent states can be
expressed as,
\begin{equation}
1 = \langle z  | z \rangle = N^2(z) \sum_{n,m = 0}^{\infty}
\frac{z^n (z^*)^m}{q^{n+m} n! m!} \langle n | m \rangle
\end{equation}
Therefore, the normalization function, $N(z)$, ought to be chosen
as,
\begin{equation}
N^{-2}(z) =   \sum_{n=0}^{\infty} \left( \frac{|z|}{q}
\right)^{2n} \frac{1}{(n!)^2} = I_0\left( \frac{2 |z|}{q}\right).
\end{equation}
Then, the normalized coherent states can be written as,
\begin{equation}
|z\rangle = \left( I_0\left( \frac{2 |z|}{q} \right)
\right)^{-\frac{1}{2}} \sum_{n=0}^{\infty} \frac{|z|^{n}}{q^n n!}
|n\rangle ,
\end{equation}
or rescaling the variable $|z|$ as $|z| \rightarrow q |z|$, it can
be re-expressed as,
\begin{equation}\label{coherent states}
|z\rangle = \left( I_0\left( 2 |z| \right) \right)^{-\frac{1}{2}}
\sum_{n=0}^{\infty} \frac{|z|^{n}}{n!} |n\rangle .
\end{equation}
In terms of the creation operator, using Eq. (\ref{coherent states
creation operator}), the coherent states can then be also written
as,
\begin{equation}
|z\rangle = \left[ I_0\left( 2 |z| \right) \right]^{-\frac{1}{2}}
I_0\left( 2\sqrt{\frac{|z| A^{\dag}}{q}} \right) |0\rangle .
\end{equation}
In the configuration space, the wave functionals corresponding to
the coherent states (\ref{coherent states}) can be expressed in
terms of the scale factor, $a$, and the variable $z$, in the form,
\begin{equation}\label{coherent states configuration}
\langle a | z \rangle = \varphi_z(a) \equiv \varphi(a,z) = \left[
I_0(2|z|) \right]^{-\frac{1}{2}} \sum_{n=0}^{\infty}
\frac{|z|^n}{n!} \phi_n(a) ,
\end{equation}
where the function $\varphi(a,z)$ has to be interpreted as a
functional of paths for the scale factor, $a(t)$, and the variable
$z$. These coherent wave functionals satisfy the boundary
conditions imposed in the previous section as they are satisfied
by the Hamiltonian eigenfunctionals. When the scale factor
degenerates, in the limit $a \rightarrow 0$, using the asymptotic
expansions for Bessel's functions, the coherent wavefunctionals
can be written as,
\begin{equation}\label{limit a goes 0}
\varphi(z,a) \approx \frac{a^{\alpha}}{\sqrt{I_0(2|z|)}}
\sum_{n=0}^{\infty} \frac{|z|^n}{n!} \frac{\left( \lambda
a^q\right)^n }{2^n n!} = \frac{a^{\alpha} \, I_0\left(\sqrt{2
\lambda |z| a^q} \right)}{\sqrt{I_0(2|z|)}} ,
\end{equation}
which may express the known boundary condition of the universe. If
$\alpha$ would vanish, then, Eq. (\ref{limit a goes 0}) expressed
the Vilenkin's tunneling condition \cite{Vilenkin86} as it took on
a constant value. If $\alpha >0$, then Eq. (\ref{limit a goes 0})
would vanish in accordance with the Hartle-Hawking no boundary
proposal \cite{Hartle83}.

\begin{figure}[h]

\begin{center}

\includegraphics[width=5cm]{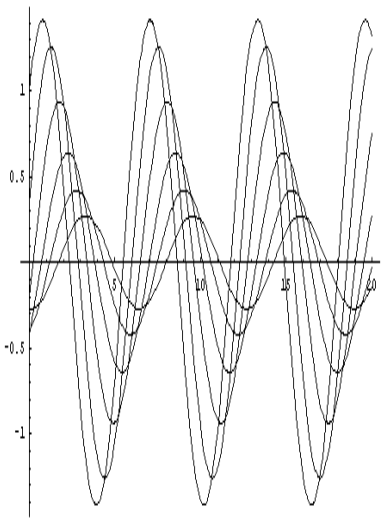}

\end{center}

\caption{Numerator in Eq. (\ref{coherent wavefunctionals action})
for different real values of the parameter $z$ (0,1,...).}

\label{univ}

\end{figure}

\begin{figure}[h]

\begin{center}

\includegraphics[width=5cm]{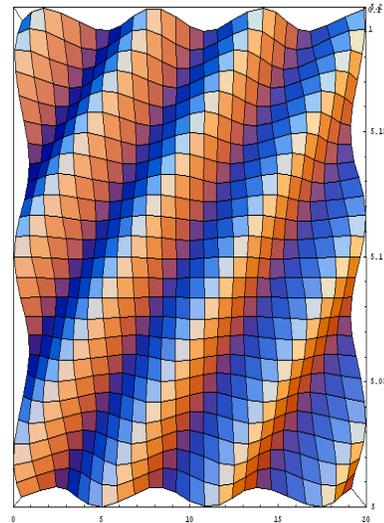}

\end{center}

\caption{Coherent wave functionals, $\varphi(z,a)$, Eq.
(\ref{coherent wavefunctionals}). It appears a set of maxima given
by $|z_k| = S_c(a) - \arctan\left( \frac{2 S_c(a) - 1}{2 S_c(a) +
1}\right) + 2 k \pi $.}

\label{univ2}

\end{figure}

In the opposite limit, for large values of the scale factor, the
introduced boundary condition is also obeyed. The limit of large
values of the scale factor is equivalent to the semiclassical
limit, where $\hbar \rightarrow 0$. In both cases, the asymptotic
expansions of Bessel's functions are the same, and the Hamiltonian
eigenfunctionals go as,
\begin{equation}
\phi_n(a) \approx \sqrt{\frac{2}{\pi \lambda a^q}} \cos\left(
\lambda a^q - \frac{\pi}{2} n - \frac{\pi}{4} \right) .
\end{equation}
Then, the coherent states can be written as,
\begin{widetext}
\begin{equation}\label{coherent wavefunctionals}
\varphi (z,a)  \approx  \frac{1}{\sqrt{I_0(2|z|)}}
\sqrt{\frac{2}{\pi \lambda a^q}} \sum_{n=0}^{\infty}
\frac{|z|^n}{n!} \cos\left( \lambda a^q - \frac{\pi}{2} n -
\frac{\pi}{4} \right) = \frac{\cos(|z|- \lambda a^q) - \sin(|z| -
\lambda a^q)}{\sqrt{\pi \lambda a^q I_0(2|z|)}} \rightarrow 0 ,
\end{equation}
\end{widetext}
for large values of the scale factor. Since in this model the
classical action is $S_c = \lambda a^q$, it turns out that the
functional $\varphi(z,a)$ can be also expressed as,
\begin{equation}\label{coherent wavefunctionals action}
\varphi (z,a)  \approx  \frac{\cos(|z|- S_c(a) ) - \sin(|z| -
S_c(a))}{\sqrt{\pi S_c(a) I_0(2|z|)}} \rightarrow 0 \; (a
\rightarrow \infty).
\end{equation}
The coherent states, in the semiclassical limit are those
represented in Fig. \ref{univ} and Fig. \ref{univ2}. There is a
set of maxima for the coherent wave functionals, for the values
$z_k$ given by,
\begin{equation}
|z_k| = S_c(a) - \arctan\left( \frac{2 S_c(a) - 1}{2 S_c(a) +
1}\right) + 2 k \pi .
\end{equation}

Therefore, we have obtained expressions for normalized coherent
states, in the configuration space. They satisfy the imposed
boundary conditions, both, in the limit of large values of the
scale factor and when it degenerates. The same limit for large
values of the scale factor runs for the semiclassical limit, in
which the coherent states should represent, by the Hartle
criterion, valid semiclassical approximations. That is the case
because Eq. (\ref{coherent wavefunctionals}) is, for any value of
the parameter $|z|$, an oscillatory function of the classical
action with a prefactor which goes to zero as the scale factor
grows up.

The second condition for coherent states to be a set of KCS
amounts to the continuity in the label $z$. It is easy to check
that this condition is satisfied. For a given pair of complex
numbers, $z = r e^{i \theta}$ and $z' = r' e^{i\theta'}$, which
are very close to one another, $r \approx r'$ and $\theta-\theta'
\approx 0$, the scalar product between them is given by,
\begin{eqnarray}\nonumber
\langle z|z'\rangle &\approx &\frac{1}{\sqrt{I_0\left(
\frac{2r}{q}\right) I_0\left( \frac{2r'}{q}\right)}}
\sum_{n=0}^{\infty} \frac{(r' r)^n}{q^{2n} (n!)^2} \\ &=&
\frac{1}{\sqrt{I_0\left( \frac{2r}{q}\right) I_0\left(
\frac{2r'}{q}\right)}} I_0\left( 2\frac{\sqrt{r' r}}{q}\right)
\approx 1 ,
\end{eqnarray}
when $r'\rightarrow r$, so the norm of the difference between two
coherent states goes to zero as they approach,
\begin{equation}
|| \, |z\rangle - |z'\rangle \, ||^2 = 2 (1-\langle z|z'\rangle)
\approx 0.
\end{equation}

The third condition on the completeness of coherent states to be a
set of KCS, can be fulfilled by including an appropriate weight
function for the integration in the variable $z$; i.e., it should
be satisfied that
\begin{equation}
\int d^2z \, W_2(z) |z\rangle \langle z | = 1 ,
\end{equation}
which in our case implies,
\begin{equation}
2 \pi \sum_{n=0}^{\infty} \frac{|n\rangle \langle n |}{(n!)^2}
\int_{0}^{\infty} d|z| \, W_2(z) \frac{|z|^{2n}}{I_0(2|z|)} = 1 .
\end{equation}
This corresponds to choosing a weight function
\begin{equation}
W_2(|z|) = \frac{2 |z|}{\pi} I_0(2|z|) K_0(2|z|) ,
\end{equation}
in the formalism of ref. \cite{Hassouni04}, and also amounts to
the fulfillment of the completeness condition. The latter
condition comes from the equalities,
\begin{eqnarray}\nonumber
\int d^2z \, W_2(z) |z\rangle \langle z | &=& 4
\sum_{n=0}^{\infty} \frac{|n\rangle \langle n |}{(n!)^2}
\int_{0}^{\infty} d|z| \, K_0(2|z|) |z|^{2n+1} \\ &=&
\sum_{n=0}^{\infty} |n\rangle \langle n | = 1 ,
\end{eqnarray}
where we have used \cite{Gradshteyn03},
\begin{equation}
\int_0^{\infty} dx \, K_0(2x) x^{2n+1} = \frac{(n!)^2}{4} .
\end{equation}

\section{Conclusions and further comments}

We have obtained a set of Klauder coherent states for a dark
energy dominated universe. They satisfy the boundary conditions
and may lead to valid semiclassical approximations. Coherent
states might represent a continuous set of states ascribable to
classical universes, which are in this way interpretable as a
multiverse. The different universes differ from one another in a
smooth way by the value taken on by the parameter $z$.

The distinction between on-shell ($\hat{H}\phi_n=0$) and off-shell
($\hat{H}\phi_n \neq 0$) contributions depends on the choice of
the factor ordering. This ultimately implies that, if the factor
ordering choice becomes eventually related to the particular
choice of a time variable, the different Hamiltonian eigenstates
may represent the so-called ground state wave functional for
particular choices of the time variable, in the configuration
space. In that case, coherent states can be interpreted as the
ground state for a given time variable, i.e., for particular
reference system.

\acknowledgements The authors thank J. P. Gazeau for useful
comments and explanations. This paper was supported by CAICYT
under Research Project Nº FIS2005-01181, and by the Research
Cooperation Project CSIC-CNRST.

\end{document}